\documentclass{nature}
\usepackage[utf8]{inputenc}
\usepackage{bm}
\usepackage{amsmath}
\usepackage{amssymb}

\newcommand\subfig[2]{{Fig.~\ref{#1}{#2}}}
\newcommand\fig[1]{{Fig.~\ref{#1}}}
\renewcommand\vec[1]{\bm{#1}}
\newcommand\local{_{\text{loc}}}

\usepackage{graphicx}
\makeatletter
\let\saved@includegraphics\includegraphics
\AtBeginDocument{\let\includegraphics\saved@includegraphics}
\renewenvironment*{figure}{\@float{figure}}{\end@float}
\makeatother

\title{Thermodynamic phases in two-dimensional active matter}

\author{Juliane U. Klamser$^{1,4}$, Sebastian C. Kapfer$^2$ \& Werner 
Krauth$^{1,3,4}$}

\begin{document}

\maketitle

\begin{affiliations}
 \item Laboratoire de Physique Statistique, Département de physique de l'ENS, 
Ecole Normale Supérieure, PSL Research University, Université Paris Diderot, 
Sorbonne Paris Cité, Sorbonne Universités, UPMC Univ. Paris 06, CNRS, 75005 
Paris, France 
 \item Theoretische Physik 1, FAU Erlangen-Nürnberg, Staudtstr. 7, 91058 
Erlangen, Germany
 \item Department of Physics, Graduate School of Science, The University of 
Tokyo, 7-3-1 Hongo, Bunkyo, Tokyo, Japan
\item Max-Planck-Institut für Physik komplexer Systeme, Nöthnitzer Str. 38, 
01187 Dresden, Germany
\end{affiliations}

\begin{abstract}
Active matter has been intensely studied for its wealth of intriguing
properties such as collective motion\cite{Vicsek}, motility-induced
phase separation (MIPS)\cite{MIPS}, and giant fluctuations away from
criticality\cite{GiantNumberFluctuations}. However, the precise connection
of active materials with their equilibrium counterparts has remained
unclear.  For two-dimensional (2D) systems, this is also because the
experimental\cite{ZahnMaret1999,KeimMaret2010,ThorneyworkDullens}
and theoretical%
\cite{HalperinNelson1978,LinTrimper2006,BernardKrauth2011,KapferKrauth2015}
understanding of the liquid, hexatic, and solid equilibrium  phases and their
phase transitions is very recent. Here, we use self-propelled particles with
inverse-power-law repulsions (but without alignment interactions) as a minimal 
model
for 2D active materials.  A kinetic Monte Carlo (MC) algorithm allows us to
map out the complete quantitative phase diagram. We demonstrate that the active
system preserves all equilibrium phases, and that phase transitions are shifted
to higher densities as a function of activity. The two-step melting scenario
is maintained. At high activity, a critical point opens up a gas--liquid MIPS
region.  We expect that the independent appearance of two-step melting and of 
MIPS
is generic for a large class of two-dimensional active systems.
\end{abstract}

\begin{figure}
  \begin{center}
       \includegraphics[scale=0.73]{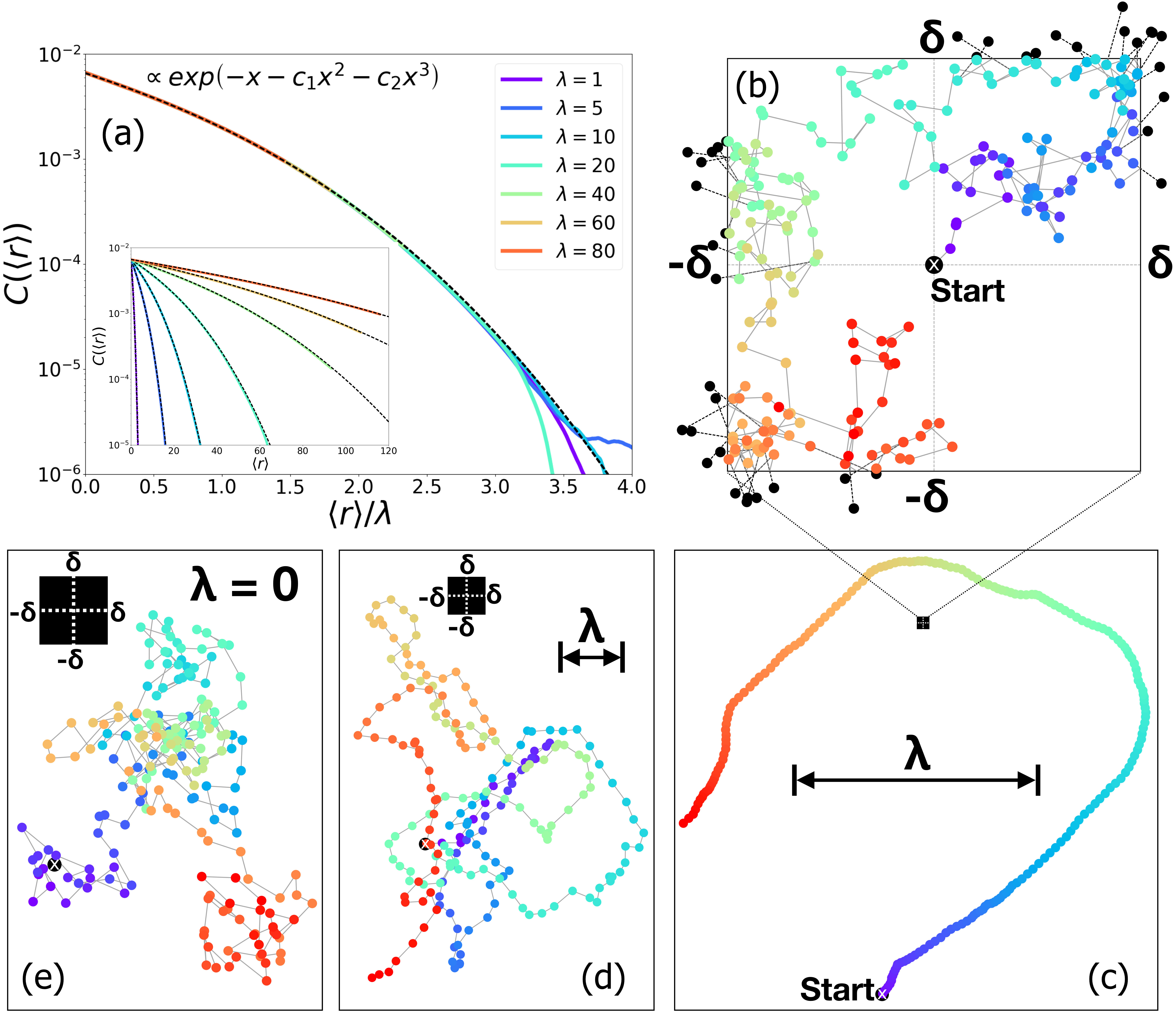}
  \end{center}
\caption{{\bf Kinetic MC algorithm.}
{\bf (a):} Autocorrelation of proposed displacements as a function of the covered distance for a
single 2D active particle.  The data collapse for widely different activities
allows the definition of a persistence length $\lambda$ (see Methods) (See
inset for raw data without rescaling).
{\bf (b):} Time evolution of proposed displacements in a box of size $[-\delta,\delta]^2$. The
displacement $\bm{\epsilon}(t)$ is sampled from a bivariate normal distribution
with standard deviation $\sigma$ (here $\sigma\ll\delta$),
centred at the previous displacement $\bm{\epsilon}({t-1})$. Positions sampled
outside the box (black points) are folded back, implementing the reflecting
boundary condition (see Methods).
{\bf (c):} Trajectory for a single particle $\bm{r}(t)
= \sum_{k = 1}^t \bm{\epsilon}(k)$, with the corresponding 
history of displacements from (b). The color gradient changing with time allows 
to connect (b) and (c), e.\,g.\ the last displacements in (b) are in the third 
quadrant thus the particle in (c) moves to the lower left etc.
{\bf (c), (d) and (e):} Sampled trajectories, illustrating the transition from 
a 
passive random walk ($\sigma \gg\delta$ in (e)) to  a persistent random walk 
($\sigma \ll \delta$ in (c)).
\label{Algorithm}
}
                
\end{figure}

The kinetic discrete-step MC dynamics for active matter progresses through 
individual particle
displacements which are accepted or rejected with the standard Metropolis
criterion. The proposed displacements of a single particle are correlated in time,
leading to ballistic local motion characterised by a persistence length
$\lambda$ that measures the degree of activity (see \fig{Algorithm}).  
Correlations decay exponentially, so that
the long-time dynamics remains diffusive. Detailed balance is satisfied only 
for vanishing activities, that is, in the
limit $\lambda \to 0$.  Cooperative effects are introduced through a repulsive 
inverse-power-law pair
potential (a $1/r^6$ potential is used). The Metropolis rejections are without
incidence on the sequence of proposed moves, which are also uncorrelated between
particles so that the active many-particle system is without alignment 
interactions.

\begin{figure}
  \begin{center}
       \includegraphics[scale=1.1]{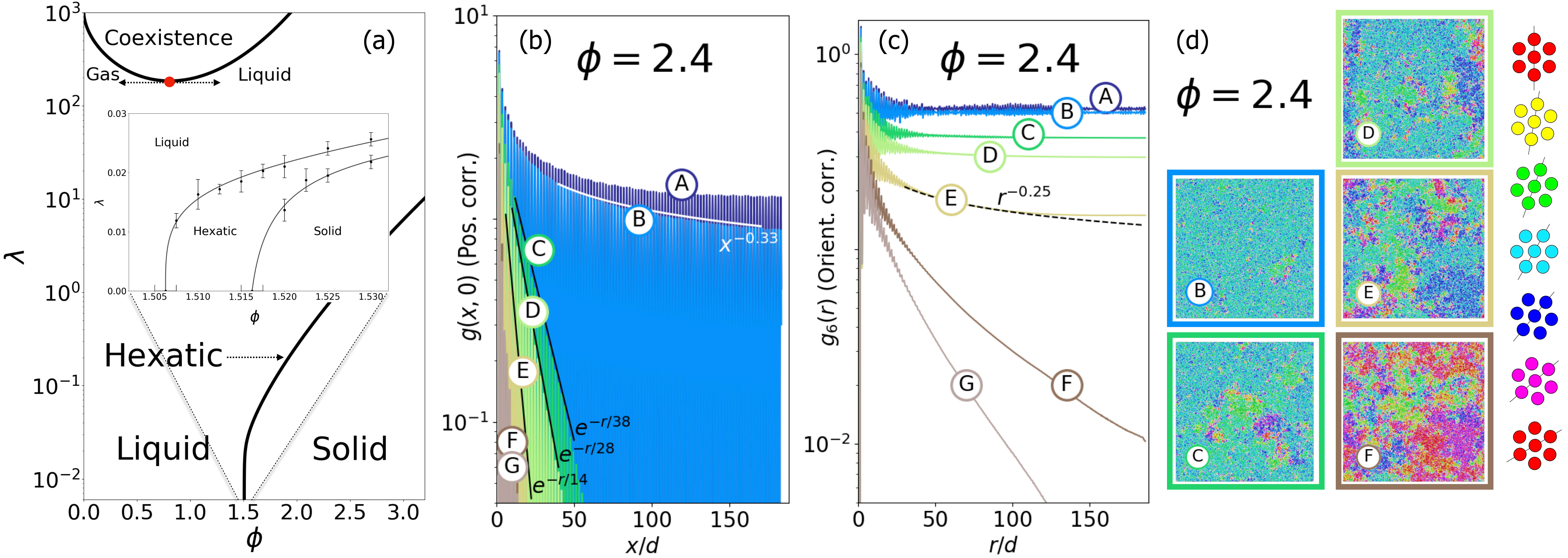}
  \end{center}
\caption{{\bf Complete phase diagram and two-step melting}. 
Depicted results are for $N \sim 4.4 \times 10^4$ particles and $\delta
= 0.1\gamma$, with $\gamma$ the particle diameter (see Methods).  
{\bf (a):} Activity $\lambda$ vs. density $\phi$ phase diagram. MIPS
between a liquid and a gas, at high $\lambda$, is situated far above the
solid--hexatic-melting lines. The red dot indicates a possible critical point.
Inset in (a): Two-step melting for small $\lambda$
with shift of transition densities to higher values with increasing
$\lambda$, preserving the equilibrium phases.  Two-step melting from the solid
is induced by density reduction (as in equilibrium) but also by an increase in
$\lambda$.
{\bf (b), (c), and (d):} Activity-induced two-step melting high above the equilibrium
melting densities ($\phi = 2.4$, A: $\lambda = 0.991\gamma$; B: $\lambda=
1.006\gamma$; C: $\lambda = 1.022\gamma$; D: $\lambda = 1.033\gamma$;
E: $\lambda = 1.049\gamma$; F: $\lambda = 1.064\gamma$; G: $\lambda =
1.079\gamma$).  {\bf (b):} Positional correlation function $g(x,y)$ along the $x$
axis, in units of the interparticle distance $d$. {\bf (c):}
Orientational correlation function $g_6(r)$ along the $x$ axis.  {\bf (d):} Snapshots
 of configurations, particles colour-coded with their local orientation
parameter $\psi_6$.  A and B are quasi-long-ranged in $g(x,0)$ and long-ranged in
$g_6$, thus corresponding to the solid phase.  C, D, and E are short-ranged
in $g(x,0)$ and quasi-long-ranged in $g_6$, thus corresponding to the hexatic
phase.  F and G decay exponentially in both correlation functions and thus
correspond to a liquid.
\label{PhaseDiagram}
}
                
\end{figure}

We observe at all densities and activities a nonequilibrium steady
state in which spatial correlation functions are well-defined. At
vanishing activity $\lambda$, we recover the equilibrium phase diagram
(see \fig{PhaseDiagram}). In particular we observe the exponential decay
of positional and orientational correlation functions in the liquid,
the power-law decay of orientational correlations yet exponential decay
of positional correlations in the hexatic phase, and long-range
orientational correlations together with positional power-law decay in the
solid.  The $1/r^6$ system is particularly amenable to simulation because
of its ``soft'' hexatic phase, characterised by small positional correlation
lengths\cite{KapferKrauth2015}.  Soft hexatics have much shorter MC
correlation times and reach the thermodynamic limit for smaller system
sizes than the hard-disk-like hexatics\cite{BernardKrauth2011} (that correspond 
to a $1 / r^n$ interaction with $n \to \infty$).   

In 2D equilibrium systems, a true crystal with Bragg peaks and long-range 
positional order exists only in the $\phi \to \infty$ limit, as a consequence of 
the Mermin--Wagner theorem\cite{MerminWagner}.  We find that
the solid phase of the active system also exhibits algebraic positional
order, just as in equilibrium. 
Decreasing the density or, remarkably, 
increasing the activity weakens 
positional correlations and eventually melts the 
solid (see~\subfig{PhaseDiagram}{a}).  Close to the melting transition,
the algebraic decay of the positional correlations can be clearly observed in
our simulation data (points A and B in \subfig{PhaseDiagram}{b}). Together with
the long-range orientational order, this explicitly identifies the solid phase
(\subfig{PhaseDiagram}{c}).

The hexatic phase in equilibrium is characterised through a lower degree
of order than the solid, namely through short-range positional correlations
(no order) and algebraic orientational correlations (quasi-long-range order).
We find precisely such a phase in the active system, in a narrow strip 
of densities below and 
activities above the solid phase (see~\subfig{PhaseDiagram}{a}).  
Starting from the solid, we indeed observe positional correlations that
change qualitatively upon a minute increase in activity while leaving the 
orientational correlations almost unchanged, leading to hexatic 
order (from point B to point C in \subfig{PhaseDiagram}{b}). 
Positional correlations in the hexatic decay exponentially beyond the 
correlation length but the orientational correlations remain quasi-long-ranged 
(points C through E in \subfig{PhaseDiagram}{c and d}).
On moving towards the liquid
at any point within the hexatic phase, the positional correlation 
lengths decrease,
resulting in a strikingly soft hexatic close to the liquid--hexatic 
transition (point E in \subfig{PhaseDiagram}{b-d}).
This soft hexatic maintains orientational quasi-order with extremely
short-ranged positional correlations, even at densities for which the equilibrium
system is already deep inside the solid phase. Increasing the activity thus
shifts the equilibrium phase boundaries towards higher densities. The stability
of the partially ordered hexatic phase is remarkable especially as it takes 
place for persistence lengths $\lambda$
significantly larger than the interparticle distance $d = (\pi N /V)^{-1/2}$.
Our massive computations give no indications of a direct transition from the 
solid into the liquid state, even at the highest accessed densities.

\begin{figure}
  \begin{center}
       \includegraphics[scale=1.05]{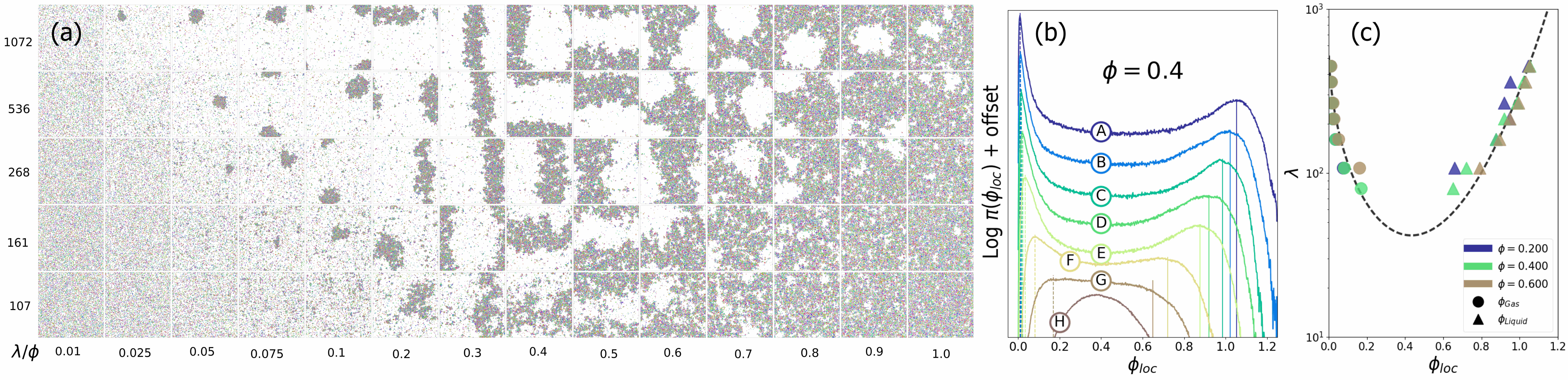}
  \end{center}
\caption{{\bf Characterisation of MIPS}. 
Data for $N \approx 1.1 \times 10^4$, $\delta = 0.7\gamma$.
{\bf (a):} Snapshots of configurations close to the onset of liquid--gas 
coexistence.
(Particles represented in arbitrary size and colour-coded, as in 
\subfig{PhaseDiagram}{d}, according to their local
orientation parameter.)  U-shaped phase boundary is apparent.  Orientational
order is short-ranged in both liquid and gas phase.  At constant activity $\lambda$,
the liquid volume fraction grows with increasing density until the liquid
entirely fills the system. The location of the critical point depends on 
$\delta$ and it appears at a
smaller density and $\lambda$ than in \fig{PhaseDiagram}.
{\bf (b):} Histograms of local densities (see Methods for definition)  for a 
variation
of activities at constant $\phi = 0.4$ (A: $\lambda = 450 \gamma$;  B: $\lambda 
= 359 \gamma$;  C: $\lambda = 268\gamma$;  D: $\lambda = 214\gamma$;  E: 
$\lambda = 161 \gamma$;  F:$\lambda = 107 \gamma$;  G: $\lambda = 80 \gamma$;  
H: $\lambda = 54\gamma$; For better presentation, histograms  are shifted along 
the $y$-axis with increasing $\lambda$).
Transition from a single-peaked to a double-peaked distribution and increasing
separation of peaks with increasing $\lambda$.
{\bf (c):} Densities of liquid and gas (identified through peak position as in (b)) in an 
activity
$\lambda$ \emph{vs.}\ local density $\phi\local$ diagram. This demonstrates
independence of phase densities on global density for fixed $\lambda$ and
validates the phase-separation picture.
\label{MIPS}
}
\end{figure}

MIPS\cite{MIPS} has been frequently reported in 2D active systems but agreement
on its interpretation was not reached. Recent work in an active
dumbbell system proposes that MIPS continuously extends from the equilibrium
liquid--hexatic transition region and that one of the separated phases
preserves some degree of order\cite{Dumbbell}. This is not the case in our
system: We observe MIPS as a U-shaped region of liquid--gas coexistence
(see \subfig{MIPS}{a}) with an onset at high activity and at relatively low density.
Both competing phases  in the phase separated state
feature exponential decay of orientational and positional
correlation functions (see the color-coded configuration snapshots in
\subfig{MIPS}{a}). MIPS is thus clearly separated from melting
(see \subfig{PhaseDiagram}{a}). In a MC simulation, a
coarsening process generally precedes macroscopic phase separation in the
time evolution towards the steady state.   In the active $1/r^6$ system, the
transient coarsening leading up to MIPS can be expected to be overcome at
earlier times than for hard disks. This makes it easier to distinguish it from 
the formation of a
steady-state gel\cite{BerthierHardSpheres}, although we do not expect the
nature of the coexisting phases participating in MIPS to depend on the softness 
of the potential.

\begin{figure}
  \begin{center}
       \includegraphics[scale=0.5]{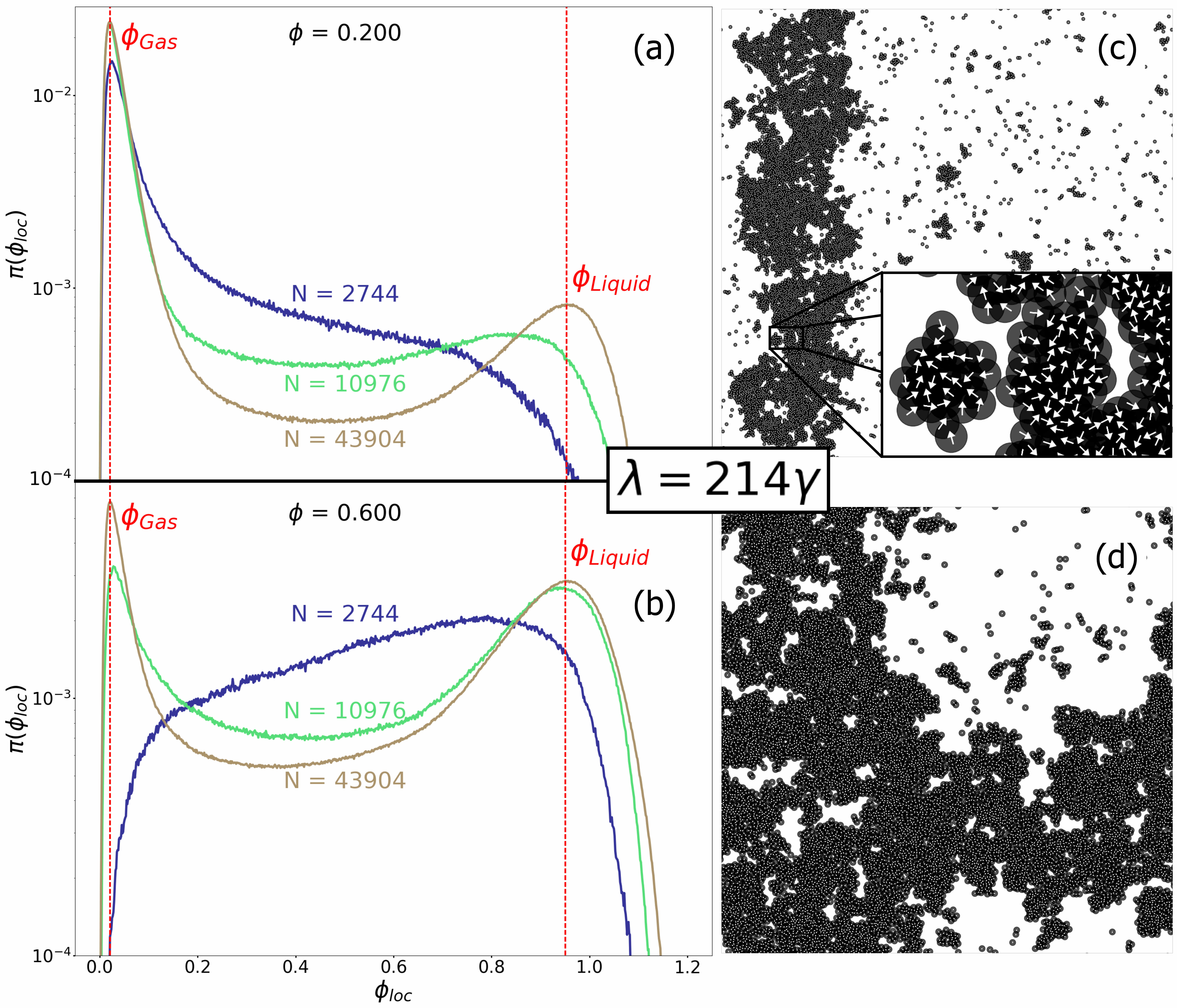}
  \end{center}
\caption{
{\bf Finite-size scaling of local density histograms} 
Data with  $\lambda=214\gamma$ and $\delta=0.7\gamma$.  
{\bf (a) and (b):}  Local density histograms (see Methods for definitions)  for 
global densities $\phi=0.2$ and $\phi=0.6$ (identical $x$-axis used). 
Local density peaks sharpen with increasing system size $N$, and are located
at the same value of $\phi\local$, demonstrating that in the MIPS region
gas and liquid densities are independent of the global density. 
{\bf (c) and (d):} Snapshots of configurations at global densities corresponding to
(a), where the liquid is the minority phase,  and (b), where the liquid is the
majority phase  by volume fraction (cf.\ height of the peaks in (a) and (b)).
{\bf Inset in (c):} Direction of motion of the individual particles
indicated by arrows, illustrating the origin of MIPS. The directions of motion 
are uncorrelated inside the homogeneous liquid and gas. Only particles at the 
interface move towards the interior of the liquid 
patch, enclosing particles of the high density region. 
\label{FiniteSizeScal}
}
                
\end{figure}

The analogy with the liquid--gas coexistence in equilibrium simple fluids
suggests the interpretation of the onset of MIPS as a critical 
point.  Indeed,
below the onset, the system remains homogeneous at large length scales with a
single-peaked local density distribution (see point H in \subfig{MIPS}{b}).
Above the onset of MIPS, the local densities develop a bimodal distribution.
The peak positions separate further as $\lambda$ increases, quantifying
the above-mentioned U-shape.  Moreover, at a given $\lambda$, the peak local
densities in the coexistence region are independent of the global density $\phi$
(see \subfig{MIPS}{c}).  This is further substantiated by a finite-size scaling analysis
at constant $\lambda$ (see \fig{FiniteSizeScal}). The phenomenology thus agrees with
that of an equilibrium phase coexistence where the relative proportions 
of the liquid and gas adapt to the global density, but where the degree of order 
of each of the phases and their densities remain unchanged.
Inside the coexistence region, at small densities, we observe an approximately
circular bubble of liquid inside the gas, followed by a stripe-shaped form that
winds around the periodic simulation box, and then followed by a bubble of gas
inside the liquid. In finite $(N,V,T)$ equilibrium systems, this complex behaviour
is brought about by the interface free energy\cite{MayerWood,SchraderBinder}
which vanishes at the critical point.
A detailed analysis of the homogeneous phases, but also of the phase-separated
systems, reveals the origin of the phase separation in the kinetic MC model.
In the bulk of the coexisting liquid and gas phases, the directions
of motion of the individual particles are uncorrelated beyond a very 
small length scale (see \subfig{FiniteSizeScal}{b}). At the liquid--gas interface, however, a majority
of the increment vectors point inwards towards the liquid phase. 
Even though a theoretical framework as reliable as statistical mechanics is currently lacking,
the effective cohesion in nonequilibrium is often attributed to the so-called swim pressure
due to active motion\cite{Brady2014,SolonCatesTailleur2015,SolonCatesTailleur2015_Nat}.

We thus find that two-step melting and MIPS are kept separated by the
homogeneous liquid phase, which is both the high-density end of MIPS and
the low-density end of the order--disorder transitions.
Intriguingly, the subtle hexatic phase survives at considerable activities.
Our massive computations allow us to reach the steady state even for strong
activities and for high densities, but only theory will be able to ascertain
the stability of the separating liquid phase and of the hexatic state
in the $\lambda,\phi\to\infty$ limit.  
Another open question is how the KTHNY theory of 2D melting%
\cite{HalperinNelson1978,NelsonHalperin1979,Young1979}, built for equilibrium
systems, can be extended to active systems. At the phase transitions, we
find the exponents $\simeq 0.33 $ and $\simeq 0.25$ predicted by the KTHNY theory 
for the positional and orientational correlations respectively (e.\,g.\ see 
\subfig{PhaseDiagram}{b} and {c}).  This suggests the existence of a 
coarse-grained functional that plays the role of an equilibrium free energy.  More
specifically, increased activity appears to reduce a state variable in our
system that corresponds to the pressure at equilibrium.
This interpretation reconciles both the fact
that increased activity induces melting through reduction of the effective
pressure, but also that liquid--gas phase coexistence is possible only at high activities,
i.\,e., at low effective pressure, in striking analogy to the behavior of simple fluids
in equilibrium. Further work is required to test this hypothesis.

In equilibrium, the nature of the two-step melting phase transitions depends on 
the softness of 
the particles, which can be tuned via the exponent $n$ of the inverse-power-law pair 
potential $U(r) \propto 1/r^n$. The two-step melting scenario for 
very soft particles ($n \lesssim 6$) comprises two continuous transitions, 
whereas for harder particles with $n \gtrsim 6$ the liquid--hexatic transition 
becomes first order\cite{KapferKrauth2015}.
One may speculate that the activity plays a similar role as the 
hardness of the particles and that at high activities the liquid--hexatic
transition becomes first order.

The liquid--gas coexistence phase we observe in kinetic MC is an example of
MIPS seen earlier in  active-matter models using Brownian and molecular
dynamics simulations%
\cite{CatesReviw2012,KineticModel,FlyMarchetti2012,CahnHilliard,LoewenCrystalization}. 
As we show, MIPS can be reproduced
within  kinetic MC dynamics, without added equilibrium-like mixing
steps\cite{BerthierHardSpheres}.  Indeed, the direction of the individual
persistent particle motion suffices to produce the effect: Particles may
be kinetically arrested for persistence lengths larger than the mean free
path, leading to density inhomogeneities, where particles in dense regions are walled 
in by particles coming from less dense regions.  However, this does not
unhinge the coarsening mechanism: The size of the inhomogeneous regions
increases with time, and the infinite-time steady state in a finite
system is characterized by only two coexisting regions so that, for the studied
$1/r^6$ potential, a gel phase is clearly absent. 
MIPS appears naturally in kinetic MC, and we  suggests that it is 
a generic feature of active matter in 2D and in higher dimensions. 
Our inverse-power-law interactions provide a tunable set of active-matter models
to study phase transitions and phase coexistence.

\begin{methods}
\subsection{Kinetic MC Dynamics for active particles.}
We use a modified Metropolis algorithm that breaks detailed balance. 
In each MC step, a displacement by an amount $\bm{\epsilon}_i$ is 
proposed for a single randomly chosen particle $i$. If the displacement were
implemented, it would cause a change in the total energy of the system $E\to 
E'$. The move is accepted with the Metropolis probability $P = 
\min\left[1,e^{-(E'-E)}\right]$. The total energy is given by $E=\sum_{i<j} 
U(\vec r_i-\vec r_j)$, where the power--law potential $U(r)=(\gamma/r)^6$ does 
not introduce an energy scale separate from density.
Thus, in equilibrium, the only relevant scale is the 
dimensionless density $\phi = \gamma^2 N/V$, with the effective 
temperature in the Metropolis probability scaling as $\phi^3$. 
In nonequilibrium, activity introduces a new control 
parameter, expressed in terms of the persistence length $\lambda$.
Simulations are performed in a $(7:4\sqrt{3})$ simulation box with
periodic boundary conditions and the soft-sphere potential is truncated
as follows: $\tilde U(r) = U(\min(r,1.8\gamma))$.

We introduce activity into the dynamics by choosing the proposed displacement 
$\bm{\epsilon}'_i$ of particle $i$ based on the previously proposed 
displacement $\bm{\epsilon}_i$ of the same particle.
The correlation is introduced in two stages. First, a random vector 
${\bm{\eta}}$ is sampled from a bivariate normal distribution $\propto\exp[-( 
\bm{\eta} - \bm{\epsilon}_i)^2/2\sigma^2]$, where $\sigma$ is the 
standard deviation. In the second stage, $\bm{\epsilon}'_i$ is generated from 
${\bm{\eta}}$ using the folding scheme
        \begin{equation*}
        \epsilon'_{i,z} \to  
            \begin{cases}
                q_{i,z}  - \delta         & \text{if } q_{i,z} < 2\delta    \\
                3\delta - q_{i,z}         & \text{if } q_{i,z} \ge 2\delta  \,,
            \end{cases}
        \end{equation*}
with $z \in \{\mathrm x,\mathrm y\}$ and $q_{i,z} =(\eta_z + \delta)
\operatorname{mod} 4\delta$, with
$a \operatorname{mod} b = a - b \left\lfloor \tfrac{a}{b} \right\rfloor$, i.\,e.,
$0\leq q_{i,z}<4\delta$, where $\lfloor a \rfloor$ is the floor function.
The folding scheme limits the size of the proposed displacements and keeps the 
dynamics local. The folding scheme is equivalent
to a random walk of the displacement variables $\vec\epsilon_i$ in
a $[-\delta,\delta]^2$ box with reflecting boundary conditions, see
\subfig{Algorithm}{b}.  Note that the random walk of the displacements is
independent of the positions of the particles, as the new increment persists
whether the displacement was accepted or not.

The square-shaped displacement box introduces a small degree of anisotropy
into the dynamics for $\lambda > 0$.  However, we explicitly verify that the 
resulting steady state is unaffected with respect to the properties concerning this letter.
At small densities in the gaseous state, where $\lambda$ is much 
smaller than the mean free path, the kinetic MC dynamics effectively reverts to 
detailed-balance dynamics as interactions between particles are rare. At higher 
densities, all large proposed displacements have vanishing Metropolis 
probabilities, thus leading to effectively isotropic dynamics.
In our numerical observation anisotropic effects are undetectable within other sources
of noise.

\subsection{Probability distribution of increments and persistence length.}
In continuous time, the increment variable $\vec\epsilon$ evolves according
to a diffusion equation $\partial_t P(\vec\epsilon ,t)=\tfrac{\sigma^2}{2}
\Delta_{\vec\epsilon} P(\vec\epsilon, t)$, with vanishing probability
flux through the boundary of the displacement box.
The steady-state distribution in the infinite-time limit is the uniform distribution
$P=(2\delta)^{-2}$.
By a Fourier ansatz, one readily finds that the autocorrelation time $\tau$ 
of increments is dominated by the first harmonic of the displacement
box, an that for large $t$, the autocorrelation function decays as
\begin{align}
    C(t) \equiv \langle\vec\epsilon(t_0+t)\cdot\vec\epsilon(t_0)\rangle 
        \propto
    \mathrm e^{-t/\tau}\,, \text{ where } \tau = \frac{8\delta^2}{\pi^2\sigma^2}\,.
\label{EQcorrTime}
\end{align}
The position of a free particle evolves as $\vec r(t) \equiv \vec r(t_0) + \int_{t_0}^t\,\mathrm ds\,\vec\epsilon(s)$.
For times shorter than the autocorrelation time, $t\lesssim\tau$,
its mean displacement is essentially given by the increment at time $t_0$,
\begin{align}
    \langle |\vec r(t)-\vec r(t_0)| \rangle = v\,t + \mathcal{O}(\sqrt{t}),
\label{EQMeanAbsDisp}
\end{align}
where the drift velocity $v\equiv|\vec\epsilon(t_0)|$ is given by the initial condition
of the increment, and the subleading term contains contributions by diffusion
of $\vec\epsilon$, including reflections, in the displacement box.
Averaging over all initial conditions $\vec\epsilon(t_0)$
with the steady-state uniform distribution, we obtain the mean drift velocity
\begin{equation*}
    \overline v
=
    \frac{\delta}{3}  \left[\sqrt{2}+\sinh ^{-1}(1)\right]
    \,.
\end{equation*}
Considering Eqs.~\eqref{EQcorrTime} and \eqref{EQMeanAbsDisp}, we may identify the
persistence length $\lambda \equiv \tau\,\overline v \simeq 0.62 \delta^3 \sigma^{-2}$.
The persistence length $\lambda$ offers a length scale which separates 
ballistic from diffusive motion.
The persistence length defined in this way allows to collapse data for the
increment autocorrelations
$C(\langle r\rangle = \lambda x) \propto \mathrm e^{-(x + c_1 x^2 + c_2 x^3)}$
at widely different activities (see \subfig{Algorithm}{a}).
The prefactors of the superexponential terms $c_1,\,c_2$ are obtained from a numerical fit.

\subsection{Measurements.}
The orientational order is quantified by the correlation
function \newline$g_6(r) \propto \langle \sum_{i,j}^N
\psi_6(\bm{r}_i)\psi_6(\bm{r}_j)\delta(r-r_{ij})\rangle$ of the 
bond-orientational order parameter $\psi_6(\bm{r}_i)$ calculated with Voronoi
weights\cite{VoronoiWeights}. $g_6(r)$ is a measure of the correlation of the local
sixfold orientational order at distance $r$.  Positional order is studied
with the direction-dependent pair correlation function $g(x,y)$. Before
averaging this two-dimensional histogram over configurations, each
configuration is realigned\cite{BernardKrauth2011} such that the $\Delta x$ axis points in
the direction of the global orientation parameter $\Psi_6 = \sum_i^N
\psi_6(\bm{r}_i)$. Correlation functions in
\fig{PhaseDiagram} are ensemble-averaged over 100 configurations of $\sim 4.4
\times 10^4$ particles, recorded after a warm-up time of $5 \times 10^6$ MC sweeps
, with each sweep containing $N$ MC steps. Configurations were recorded
in time intervals of $2.7 \times 10^4$ MC sweeps.

MIPS is quantified by histograms of local densities. We compute local
densities by covering the system with randomly placed test circles of
radius $7.5\gamma$.  The local dimensionless density of each test circle is
$\phi\local = \gamma^2 N\local  / V\local$, where $N\local$ are the number
of particle centres located within a circle of area $V\local$. The
detailed analysis (see Figs.~\ref{MIPS}, \ref{FiniteSizeScal}) of MIPS uses
$\delta = 0.7\gamma$. The larger $\delta$ shifts the liquid--gas
coexistence phase boundaries, in particular the critical point, to smaller
densities and persistence lengths, which drastically shortens the time to reach
the steady state. 
Configuration snapshots in \subfig{MIPS}{a} were taken after $1.1 \times 10^7$ 
MC sweeps. Data in \subfig{MIPS}{b}
consists of ensemble averages over 100 configurations recorded in 
time intervals of $3\times 10^4$ sweeps.
The $N \sim 4.4\times 10^4$ ($N \sim 1.1\times 10^4$, $N \sim 2.7 \times 10^3$) 
data in Figs.~\ref{FiniteSizeScal} was recorded after $5.5\times10^{6}$ 
($2.2\times10^{7}$, $1.4\times10^{7}$) sweeps. The average consists of 500 
configurations recorded in time intervals of $2.2 \times 10^3$ ($8.7 \times 10^3$, 
$1.7 \times 10^4$) sweeps.

\end{methods}

\section*{References}
\renewcommand\refname{ }


\begin{addendum}
 \item We thank Hughes Chaté and Hartmut Löwen for helpful discussions. 
 W.K. acknowledges support from the Alexander von Humboldt Foundation.
\item[Email addresses:] juliane\texttt{\char`_}klamser@yahoo.de, sebastian.kapfer@fau.de, 
werner.krauth@ens.fr.
\end{addendum}


\end{document}